\documentclass[aps,pra,showpacs,superscriptaddress,groupedaddress,twocolumn,nofootinbib]{revtex4-2}  

\usepackage[utf8]{inputenc}
\usepackage{bm}
\usepackage{color}

\usepackage{hyperref}
\usepackage{amsmath, mathtools}
\usepackage{amssymb}
\usepackage{wasysym}
\usepackage{graphicx} 
\usepackage{booktabs} 
\usepackage{pdfsync}
\usepackage{verbatim}
\usepackage{latexsym}
\usepackage{dsfont}

\usepackage{float}

\usepackage[bottom]{footmisc}
\usepackage{subcaption}

\def\sec#1{\section{#1} }
\def\ssec#1{\subsection{#1} }

\def\({\left(}
\def\){\right)}
\def\[{\left[}
\def\]{\right]}

\def\a{\alpha}

\def\f#1#2{\frac{#1}{#2}}

\def\de{\delta}

\def\l{\lambda}
\def\L{\Lambda}

\def\<{\langle}
\def\>{\rangle}

\providecommand{\abs}[1]{\left\lvert#1\right\rvert}

\usepackage{graphicx}
\usepackage{dcolumn}
\usepackage{bm}

\begin{document}
\title{Fitting for the energy levels of hydrogen}

\author{David~M.~Jacobs}
\email{david.m.jacobs@asu.edu}
\thanks{Present Address: Department of Physics, Arizona State University, Tempe, AZ 85281}
\affiliation{Physics Department, Norwich University, Northfield, Vermont 05663, USA}
\affiliation{Physics Department,
Case Western Reserve University,
Cleveland, Ohio 44106, USA}

\author{Marko Horbatsch}
\email{marko@yorku.ca}
\affiliation{Department of Physics and Astronomy, York University, Toronto, Ontario, Canada M3J 1P3}

\date{\today}


\begin{abstract}
Atomic hydrogen energy levels calculated to high precision are required to assist experimental researchers working on
spectroscopy in the pursuit of testing quantum electrodynamics (QED) and probing for physics beyond the Standard Model. There are two important parts to the problem of computing these levels: an accurate evaluation of contributions from QED and using an accurate value for the proton charge radius as an input. Recent progress on QED corrections to the fine structure, as well as increasing evidence that a proton charge radius in the range of 0.84 fm is favored over the previously adopted larger value in the 0.88 fm range, has advanced the field, yet several state-of-the-art measurements remain in contradiction with this smaller value. Motivated by on-going and future work in this area, we present here a simple parameterization for the energy levels of hydrogen at the level of hyperfine structure using the so-called relativistic Ritz approach. The fitting of a finite sample of QED-generated levels at low to intermediate principal quantum number, $n$, gives a generally applicable formula for \emph{all} values of $n$ for each distinct angular momentum channel, given in this work up to orbital angular momentum number $\ell=30$. We also provide a simple linear parameterization for the shift in hydrogen energy levels as a function of the proton radius, providing a useful cross check for extant and future measured energy intervals.
\end{abstract}
\maketitle


\sec{Introduction}\label{Sec:intro}
~

Precision measurements of atoms are used for metrological purposes and  testing the theory of quantum electrodynamics (QED). This is of current interest also in the context of beyond-the-Standard-Model phenomena, as they could manifest themselves in atomic spectroscopy \cite{safronova2018search}. The theory of bound-state QED is sufficiently mature that the dominant uncertainty in its predictions for the levels of hydrogen and deuterium is due to the nuclear radius. 
 The proton radius puzzle first appeared in 2010 \cite{Pohl:2010zza} when muonic hydrogen measurements indicated that $r_p$ is 4\% smaller than had been previously determined, a value near $0.88$ fm \cite{mohr2012codata}. Over the last decade or so, more scattering and spectroscopic experiments have been performed that suggest a value of $r_p$ closer to $0.84$ fm \cite{ubachs2020crisis}. However, discrepancies remain, such as the results of Fleurbaey et al. \cite{fleurbaey2018new} and Brandt et al. \cite{Brandt:2021yor}, that indicate a value of $r_p$ larger than $0.84$ fm with substantial statistical significance. Thus, the puzzle is not entirely solved; more measurements are planned in the near future, including that of the $1S_{1/2}\to 4S_{1/2}$ interval \cite{yzombard20231s}.

The bound-state QED predictions for the energy levels of hydrogen involve a combination of long analytic expressions and numerical results that are cumbersome to use; see, e.g., \cite{Tiesinga:2021myr}. Our goal here is to provide a fitting formula that reproduces the bound-state QED predictions for those energy levels to sufficiently high accuracy that bound-state QED need not be used directly.  To this end, we use the so-called relativistic Ritz approach, which is a long-distance effective theory describing the bound states of two-particle systems whose binding potential is dominated by the Coulomb interaction \cite{Jacobs:2022mqm}. In that effective theory, the energy levels of atomic hydrogen were shown to be
\begin{equation}\label{general_E_sol}
\f{E}{c^2}=\sqrt{m_e^2+m_p^2+\f{2m_em_p}{\sqrt{1+\(\f{\alpha}{n_\star}\)^2}}} - \Big(m_e+m_p\Big)\,,
\end{equation}
\\where $\a$ is the fine-structure constant and the effective quantum number
\begin{equation}
n_\star=n-\de\,.
\end{equation}
The quantum defect, $\de$, itself depends on the principal quantum number, $n$, and accounts for interactions that are shorter in range than the Coulomb interaction; it also depends on the orbital, total electronic, and total system quantum numbers, $\ell$, $j$, and $f$, respectively.

To make the numerical analyses more efficient, we Taylor expand equation \eqref{general_E_sol} in small $\a$ up to eighth order\footnote{There is no ninth order term so this is sufficient for the accuracy required here. Furthermore, the highest computed QED terms have so far not been computed at higher precision than this \cite{Tiesinga:2021myr}.} and factor out the Rydberg frequency,
\begin{equation}
cR_\infty\equiv\f{m_e\a^2 c^2}{2h}\,,
\end{equation}
allowing us to write
\begin{equation}\label{Rel_Ritz_series}
\f{E}{h}=cR_\infty\(\f{A_2}{n_\star^2}+\f{A_4}{n_\star^4}+\f{A_6}{n_\star^6}+\f{A_8}{n_\star^8}\)\,,
\end{equation}
where the $A_{2k} ={\cal O}\!\(\alpha^{2k-2}\)$. For the value of the fine-structure constant we use 
\begin{equation}
\a^{-1}=137.035\,999\,166(15)\,,
\end{equation}
derived from a recent measurement of the electron g-factor \cite{Fan:2022eto}. Together with the mass ratio 
\begin{equation}
\frac{m_p}{m_e}=1\,836.152\,673\,349(71)\,,
\end{equation}
inferred from spectroscopy of $\text{HD}^+$ \cite{Patra:2020brw}\footnote{{This value is consistent with that of Ref. \cite{alighanbari2020precise}. Both values reported in Refs. \cite{Patra:2020brw} and \cite{alighanbari2020precise} rely on the proton-to-deuteron mass ratio obtained by Fink and Myers \cite{fink2020deuteron}.}}, this allows us to determine the constants
\begin{eqnarray}
A_2 &=&-0.999\,455\,679\,424\,739(21) \\
A_4 &=& 3.990\,953\,7921(87)\times10^{-5} \\
A_6 &=&-1.770\,774\times 10^{-9} \\
A_8 &=& 8.25\times10^{-14}\,.
\end{eqnarray}
There are uncertainties in $A_6$ and $A_8$; however, they are irrelevant at the level of accuracy needed here.

The simplest Ritz-like expansion is posited for the quantum defect, namely a series expansion in terms of the energy eigenvalues, which are assumed to be small relative to some high-energy scale, $\Lambda$:
\begin{equation}\label{defect_ansatz}
\de_{}=\de_{0}+\l_{1} \f{E}{\L}+\l_{2}\(\f{E}{\L}\)^2+\dots\,,
\end{equation}
where $\de_{0}$ and the $\l_i$ are dimensionless coefficients. However, because in this form $E$ depends implicitly on $\de$, it is impractical to use for most theoretical or empirical applications. A \emph{modified} ansatz written as a series in inverse powers of $(n-\de_{0})$ is asymptotically ($n\to\infty$) equivalent to \eqref{defect_ansatz} and is significantly easier to use for data fitting. Analyzing the large-$n$ behavior of \eqref{Rel_Ritz_series} with \eqref{defect_ansatz}, it may be verified that

\begin{widetext}
\begin{multline}\label{defect_modified_ansatz}
\de_{}=\de_{0}+ \f{\de_{2}}{\(n-\de_{0}\)^2}+\f{\de_{4}}{\(n-\de_{0}\)^4} +\f{2\de_{2}^2}{\(n-\de_{0}\)^5}  +\f{\de_{6}}{\(n-\de_{0}\)^6}  +\f{6\de_{2} \de_{4}}{\(n-\de_{0}\)^7}+\f{\de_{8}}{\(n-\de_{0}\)^8}  +\f{4\de_{4}^2 + 8\de_{2} \de_{6}}{\(n-\de_{0}\)^9}\\
+\f{\de_{10}}{\(n-\de_{0}\)^{10}} +\f{-40\de_{2}^4 + 10\de_{4} \de_{6}+ 10\de_{2} \de_{8}}{\(n-\de_{0}\)^{11}}+\f{\de_{12}}{\(n-\de_{0}\)^{12}} +\f{-296 \de_2^3 \de_4+6\de_6^2+12\de_4\de_8+12 \de_2 \de_{10}}{\(n-\de_{0}\)^{13}}+\dots\,,
\end{multline}
\end{widetext}
where the $\de_i$ are free parameters. As shown in the following section, $\delta_0$ is small (of order $\alpha^2$) and thus $1/(n-\delta_0)$ is small for $n>1$, but we find that the modified defect expansion \eqref{defect_modified_ansatz} satisfactorily reproduces energy levels even for $n=1$ with the inclusion of a sufficient number of $\de_i$.

A truncation of equation \eqref{defect_modified_ansatz} is required for any application and we specify the order of the analysis by the highest inverse power of $(n-\de_0)$ included. Actually, truncations made at each successive inverse \emph{odd} power includes one additional defect parameter. Because there is no $-1$st or $-3$rd term, including terms through $(n-\de_0)^{-1}$ requires only $\de_0$ and is considered lowest order (LO), whereas including terms through $(n-\de_0)^{-3}$ requires both $\de_0$ and $\de_2$ and is considered next-to-lowest order (NLO). At higher orders we use the abbreviation  N$^{k}$LO, where $k+1$ is equal to the number of defect parameters needed.  For practical purposes, as shown below, the largest expansion is needed for $S$-states, where we truncate at the level N$^6$LO, thereby including defect parameters up to $\delta_{12}$. For higher angular momentum eigenstates fewer terms are required to reach the same level of accuracy.

As outlined below, we use a limited number of precisely calculated energy levels of hydrogen employing the most up-to-date bound-state QED calculations \cite{Tiesinga:2021myr}, and fit them with equation \eqref{Rel_Ritz_series} using the defect formula in equation  \eqref{defect_modified_ansatz}. We determine the necessary $\de_i$ to reproduce all theoretical energy levels to within their uncertainties and demonstrate the power of our fits by testing our results against higher-$n$ calculated energies.  Because some energies can be predicted with a relative precision that is better than $10^{-13}$, to ensure this level of reproducibility, the parameters $A_2$ through $A_8$ are given to an absolute precision of $10^{-15}$ and, likewise, we report our fit values of the $\de_i$ to the same level of precision.

\sec{Theoretical inputs, uncertainties, and shifts due to the proton radius}

According to bound-state QED, the theoretical energy levels of hydrogen can be written as the sum of a gross level structure, fine-structure (FS), and hyperfine-structure (HFS) contribution,
\begin{equation}\label{QED_E_levels}
E_{n\ell j f}= -\f{cR_\infty}{n^2}\f{m_p}{m_p+m_e} + E_{n\ell j}^{(\text{FS})} + E_{n\ell jf}^{(\text{HFS})}\,,
\end{equation}
where we have chosen units in which Planck's constant, $h=1$. The electron's reduced Compton wavelength, muon-to-electron mass ratio, proton g-factor, and electron magnetic-moment anomaly taken from CODATA-18 \cite{Tiesinga:2021myr} are
\begin{eqnarray}
\lambdabar_e &=& 3.861\,592\,6796(12)\times 10^{-13}\,\text{m}\\
\f{m_\mu}{m_e} &=& 206.768\,2830(46)\\
g_p &=& 5.585\,694\,6893(16)\\
a_e &=& 1.159\,652\,181\,28(18) \times10^{-3}\,.
\end{eqnarray}
We also use the proton radius inferred from the muonic hydrogen spectroscopy of Antognini et al. 2013 \cite{Antognini:1900ns}, 
\begin{equation}\label{r_p_Antognini}
r_p = 0.840\,87(39)\, \text{fm}\,.
\end{equation}
Following the procedure described in \cite{horbatsch2016tabulation}, the measured $1S_{1/2}$ \cite{karshenboim2005precision} and $2S_{1/2}$ \cite{kolachevsky2009measurement} hyperfine intervals may be used to the determine $E_{n\ell jf}^{(\text{HFS})}$ to sufficient accuracy such that $cR_\infty$ may be determined using the measured $1S^{(f=1)}_{1/2}\to2S^{(f=1)}_{1/2}$ interval from \cite{Parthey:2011lfa}; this completely specifies the theory.

The theoretical uncertainty in the energy levels is dominated by the uncertainty in $E_{n\ell j}^{(\text{FS})}$, which affects the levels directly through $E_{n\ell j}^{(\text{FS})}$ itself and also indirectly through the determination of $cR_\infty $. There are 5 uncertainties relevant to $E_{n\ell j}^{(\text{FS})}$ at the level of precision needed in this work. Four of these are QED uncertainties taken directly from CODATA-18 \cite{Tiesinga:2021myr}: the uncertainty in the two-photon correction term $B_{60}$ yields $\de_{\ell,0}\(0.94\,\text{kHz}\)/n^3$ (a reduction by about 50\% compared to CODATA-14); the uncertainty in the three-photon correction term $C_{50}$ yields $\de_{\ell,0}\(0.96\,\text{kHz}\)/n^3$; nuclear polarizability uncertainty yields $\de_{\ell,0}\(0.39\,\text{kHz}\)/n^3$; and a radiative recoil uncertainty yields $\de_{\ell,0}\(0.74\,\text{kHz}\)/n^3$.

In addition to the QED uncertainties mentioned above, there is an error in $E_{n\ell j}^{(\text{FS})}$ due to the proton radius \eqref{r_p_Antognini} which amounts to  $\de_{\ell,0}\(1.03\,\text{kHz}\)/n^3$. Adding all of these errors in quadrature, the overall uncertainty in the fine-structure correction is
\begin{equation}
\de(E_{n\ell j}^{(\text{FS})})=     \f{\(1.9 \,\text{kHz}\)}{n^3}\de_{\ell,0}\,,
\end{equation}
and it follows that
\begin{equation}\label{new_Rydberg}
cR_\infty = 3\,289\,84\,960\,249.1(2.2) \,\text{kHz}\,,
\end{equation}
a shift upward of 0.2 kHz compared to the result reported in \cite{horbatsch2016tabulation}, but well within the uncertainty computed therein. Accounting for the correlated uncertainties in the QED predictions and determination of $cR_\infty$, the theoretical uncertainty on any given level is 
\begin{equation}\label{QED_level_unc}
\de(E_{n\ell j f})=\abs{     \f{1.9 \,\text{kHz}}{n^3}\de_{\ell,0}    - \f{2.2 \,\text{kHz}}{n^2}     }\,,
\end{equation}
and is therefore below $0.6$ kHz for all levels. 

Lastly, given the potential issue of a remaining proton radius puzzle, we consider the possible systematic implications of a shift away from the value quoted in equation \eqref{r_p_Antognini}. Defining the proton radius shift,
\begin{equation}
\Delta r_p = r_p - 0.840\,87\,\text{fm}\,,
\end{equation}
it follows that the Rydberg frequency shifts by
\begin{equation}\label{Rydberg_change_due_to_rp}
\Delta \(cR_\infty\)_{r_p}=     3.1 \,\text{kHz}\(\f{\Delta r_p}{0.001\,\text{fm}}\)\,,
\end{equation}
and the energy levels shift by
\begin{equation}\label{Elevel_change_due_to_rp}
\Delta \(E_{n\ell jf}\)_{r_p}=   \(\f{2.6 \,\text{kHz}}{n^3}\de_{\ell,0} - \f{3.1 \,\text{kHz}}{n^2}\)\(\f{\Delta r_p}{0.001\,\text{fm}}\)\,.
\end{equation}
Equations \eqref{Rydberg_change_due_to_rp} and \eqref{Elevel_change_due_to_rp} will be utilized below to cross check our results against a selection of experimental results.

\sec{Fitting to theoretical levels of hydrogen }\label{Sec:Fitting_H}

\ssec{Overview}

For a given orbital angular momentum value, $\ell$, we generate values of $E_{n\ell j f}$ using \eqref{QED_E_levels}{, following the same procedure described in Ref. \cite{horbatsch2016tabulation} with updated theoretical inputs from Ref. \cite{Tiesinga:2021myr}.  Values of Bethe logarithms are taken from Refs. \cite{Drake:1990zz} and \cite{jentschura2005calculation}; however, many levels require numerically-computed QED terms, such as $B_{60}(n\ell_{j})$, which have not been computed (or made publicly available) for all values of $n$, $\ell$, and $j$. Therefore, we fit the available values of such terms with simple formulas in terms of inverse powers of $n$ and interpolate or extrapolate to obtain the needed terms. Our conservative estimates for the interpolation/extrapolation error is far below the theoretical (QED) error.}

\begin{figure}[!htb]
    \includegraphics[width=8.6cm]{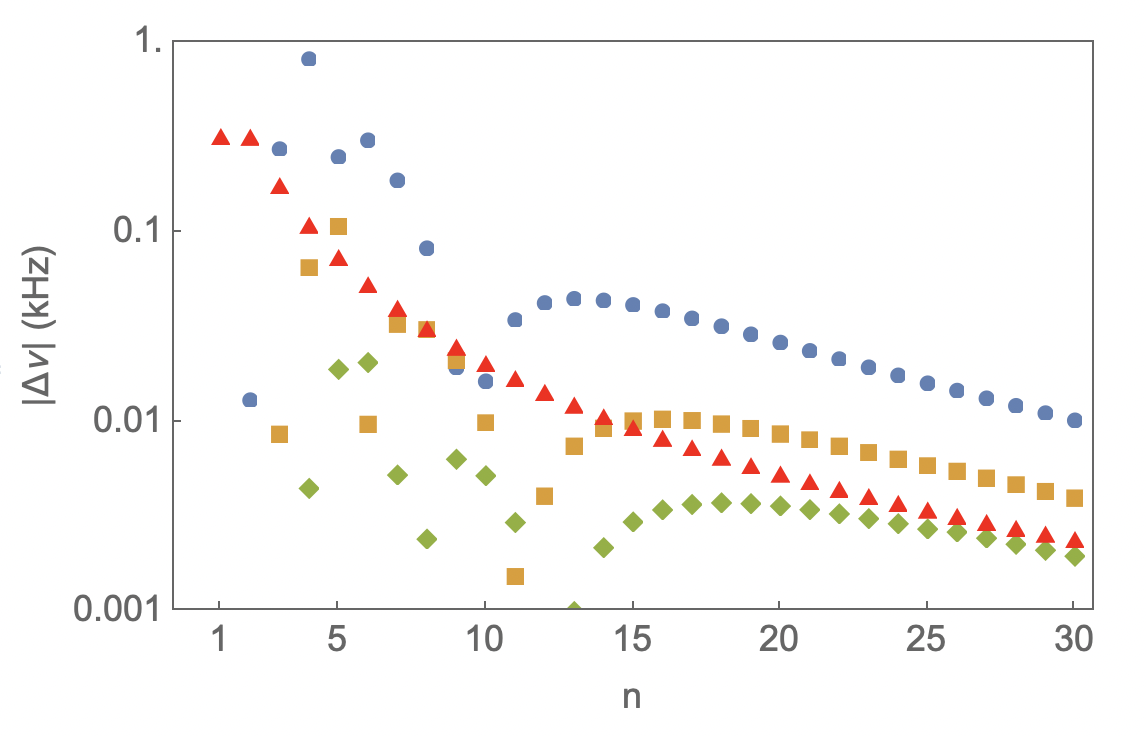}
  \caption{\raggedright Absolute differences between fit values of the energy levels from equation \eqref{Rel_Ritz_series} and QED theory from equation \eqref{QED_E_levels} for $S^{(f=1)}_{j=1/2}$ states. The $\rm N^4LO$, $\rm N^5LO$, and $\rm N^6LO$ fit differences are indicated by solid circles, squares, and diamonds, respectively, while the QED theory error is indicated with solid triangles. }
    \label{Fig:Sf1_fitting_compare.png}
  \end{figure}

\begin{figure}[!htb]
    \includegraphics[width=8.6cm]{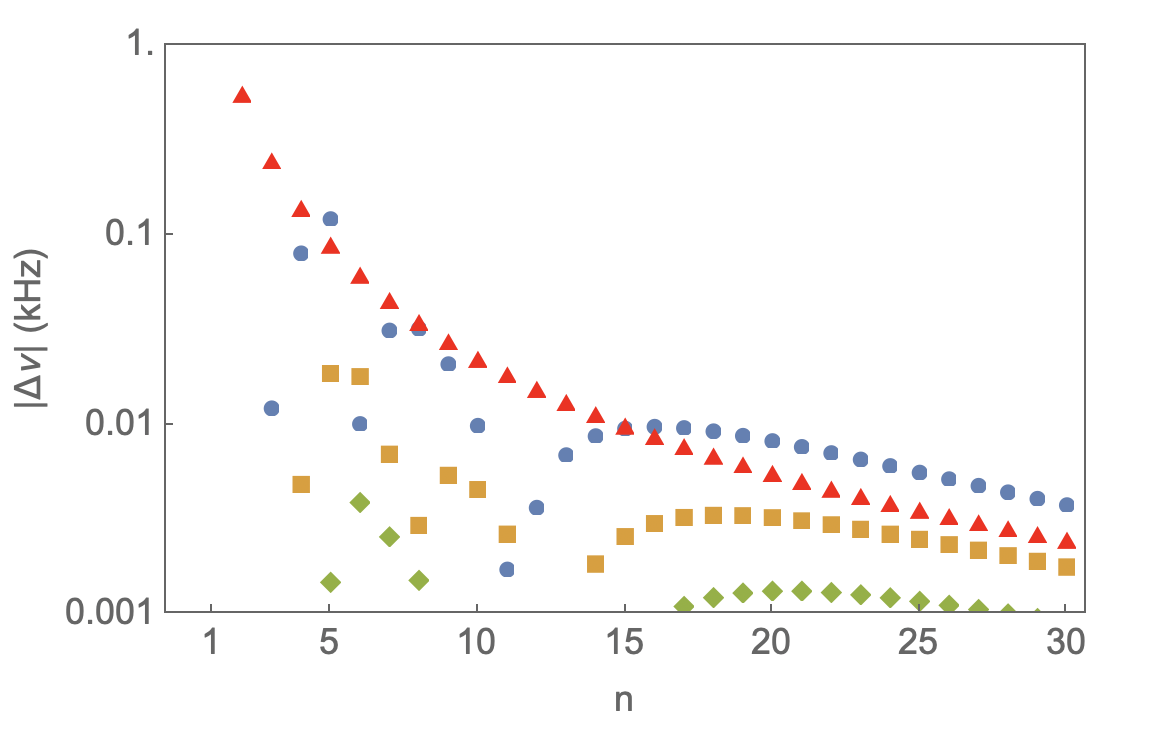}
  \caption{\raggedright Same as in Figure \ref{Fig:Sf1_fitting_compare.png} but for  $P^{(f=1)}_{j=3/2}$.}
    \label{Fig:Pf1_fitting_compare.png}
  \end{figure}

We compute energy levels from $n_\text{min}=\ell+1$ up to $n_\text{max}=\max{\(15,\ell+1\)}$, fit them with equation \eqref{Rel_Ritz_series} using the defect formula in equation  \eqref{defect_modified_ansatz}, and weight each data point by the inverse square of the theoretical uncertainty given in \eqref{QED_level_unc}. The fit order, i.e. the number of necessary defect parameters ($\de_i$), is increased until the difference between the fit value and the QED value from \eqref{QED_E_levels} falls below the QED error given in equation \eqref{QED_level_unc}. Some example fits for a subset of $\ell=0$ and $\ell=1$ states are shown in Figures \ref{Fig:Sf1_fitting_compare.png} and \ref{Fig:Pf1_fitting_compare.png}, which show the absolute differences between the relativistic Ritz and QED predictions; differences that do not appear in the figures are below $1\,\text{Hz}$. For these states, we have used QED-predicted levels up to $n_\text{max}=15$ for the fits, so all $n>15$ values represent a true test of the model against QED.

  \begin{figure}[!htb]
    \includegraphics[width=8.6cm]{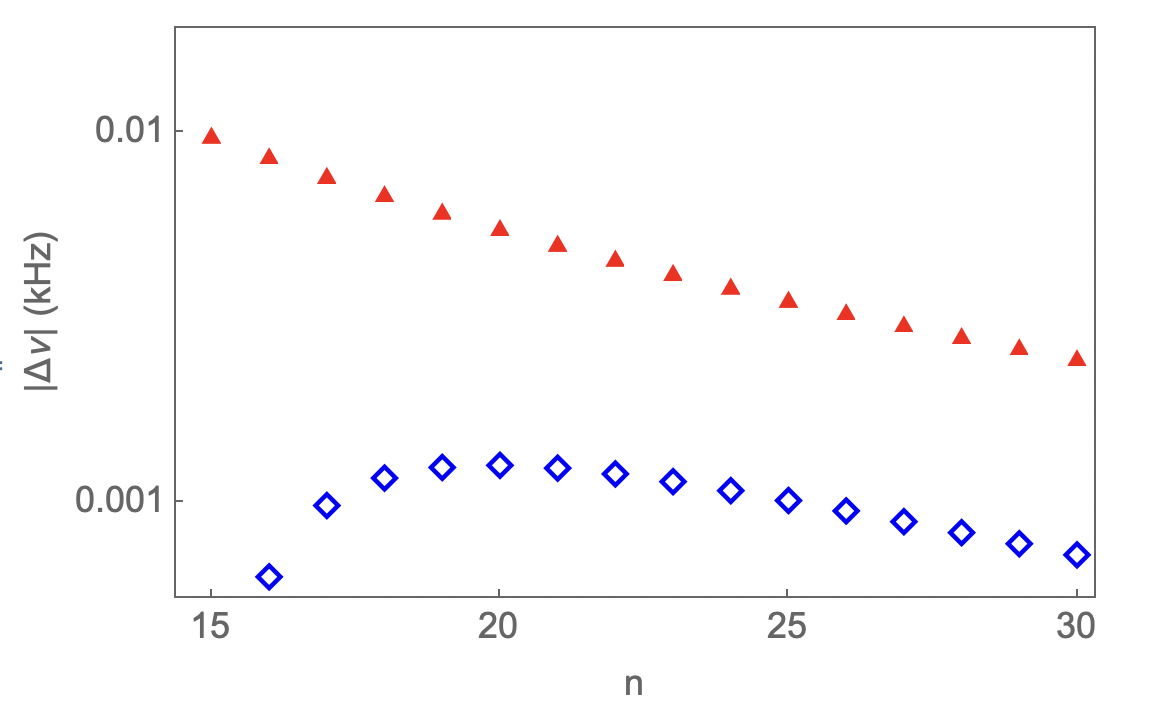}
  \caption{\raggedright Analogous to Figure \ref{Fig:Sf1_fitting_compare.png} but for  $\ell=14$, $j=29/2$, and $f=15$. Open diamonds indicate differences for the $\rm LO$ fit, while solid triangles indicate the QED theory error. Notably, only one QED-predicted level ($n=15$) was used to fit for $\delta_0$.}
    \label{Fig:l14_fitting_compare.png}
  \end{figure}

Summarizing the findings presented in Figs. \ref{Fig:Sf1_fitting_compare.png} and \ref{Fig:Pf1_fitting_compare.png} we make the following observations:
The number of fit parameters required, i.e., the order of the expansion in \eqref{defect_modified_ansatz} depends
on the angular momentum value $\ell$. This is related physically to the fact that states with $\ell>0$ have a centrifugal barrier preventing the
electron from getting close to the proton. For $S$-states the complete model as written out in \eqref{defect_modified_ansatz}
is required, and with increasing integer values of $\ell$ the required order tends to decrease until $\ell=9$, beyond which only $\delta_0$ is needed. This trend is partially demonstrated in Fig.~\ref{Fig:Sf1_fitting_compare.png}, where $\ell=0$ and the $\rm N^6LO$ model provides an adequate fit, while in Fig.~\ref{Fig:Pf1_fitting_compare.png} ($\ell=1$) the $\rm N^5LO$ model is shown to provide sufficient accuracy. In either case, the low-$n$ levels are reproduced with such precision ($<1$ Hz) that they do not appear in the figures.

 The full set of fit parameters for $\ell=0$ and $\ell=1$ states is shown in Table \ref{fitting_parameter_table_l01}, and Tables \ref{fitting_parameter_table_l234} - \ref{fitting_parameter_table_l21-30} in Appendix \ref{Appendix_Parameters} present fit parameters for states from $\ell=2$ to $\ell=30$. The number of fit parameters for each combination of $\ell, j$, and $f$ never exceeds the number of energy values used for each fit. In fact, for states with $\ell\geq14$ we have only used one QED-predicted level to fit for the one defect parameter ($\delta_{0}$) required and have verified our model predictions up to at least $n=30$; see Fig. \ref{Fig:l14_fitting_compare.png} for an example in which $\ell=14$. This points to the efficiency of the relativistic Ritz family of models.

  \begin{table}[h]
    \centering
    \begin{tabular}{c  c c c r}
\hline\hline
      \multicolumn{1}{c}
~ $\ell~$ &~$j$~&~$f$~& $\de_i$ & Value$/10^{-5}$  \\ \hline
$0$ &  $\f{1}{2}$ & $0$  & $\de_0$ & $2.550\,210\,0611$  \\
&    &    & $\de_2$ & $0.008\,375\,5621$  \\
&    &    & $\de_4$ & $-0.031\,662\,6562$  \\
&    &    & $\de_6$ & $0.271\,465\,6639$  \\
&    &    & $\de_8$ & $-1.493\,445\,3489$  \\
&    &    & $\de_{10}$ & $3.605\,726\,0079$  \\
&    &    & $\de_{12}$ & $-2.356\,090\,4800$  \\[0.1cm]
$0$ &  $\f{1}{2}$ & $1$  & $\de_0$ & $2.528\,610\,1667$  \\
&    &    & $\de_2$ & $0.008\,375\,3687$  \\
&    &    & $\de_4$ & $-0.031\,651\,9458$  \\
&    &    & $\de_6$ & $0.271\,324\,9311$  \\
&    &    & $\de_8$ & $-1.492\,530\,9405$  \\
&    &    & $\de_{10}$ & $3.603\,313\,0305$  \\
&    &    & $\de_{12}$ & $-2.354\,461\,9943$  \\[0.1cm]

$1$ &  $\f{1}{2}$ & $0$  & $\de_0$ & $2.669\,249\,8201$  \\
&    &    & $\de_2$ & $0.002\,680\,7904$  \\
&    &    & $\de_4$ & $-0.023\,483\,8927$  \\
&    &    & $\de_6$ & $0.234\,141\,1690$  \\
&    &    & $\de_8$ & $-1.259\,283\,0776$  \\
&    &    & $\de_{10}$ & $2.428\,620\,8124$  \\[0.1cm]

$1$ &  $\f{1}{2}$ & $1$  & $\de_0$ & $2.662\,051\,7448$  \\
&    &    & $\de_2$ & $0.002\,681\,1814$  \\
&    &    & $\de_4$ & $-0.023\,484\,0422$  \\
&    &    & $\de_6$ & $0.234\,143\,0659$  \\
&    &    & $\de_8$ & $-1.259\,294\,3395$  \\
&    &    & $\de_{10}$ & $2.428\,643\,5695$  \\[0.1cm]

$1$ &  $\f{3}{2}$ & $1$  & $\de_0$ & $1.331\,250\,5239$  \\
&    &    & $\de_2$ & $0.002\,680\,3758$  \\
&    &    & $\de_4$ & $-0.023\,507\,0995$  \\
&    &    & $\de_6$ & $0.234\,315\,4776$  \\
&    &    & $\de_8$ & $-1.259\,730\,2288$  \\
&    &    & $\de_{10}$ & $2.428\,830\,0568$  \\[0.1cm]

$1$ &  $\f{3}{2}$ & $2$  & $\de_0$ & $1.328\,373\,2345$  \\
&    &    & $\de_2$ & $0.002\,680\,4364$  \\
&    &    & $\de_4$ & $-0.023\,507\,0852$  \\
&    &    & $\de_6$ & $0.234\,315\,3010$  \\
&    &    & $\de_8$ & $-1.259\,729\,2077$  \\
&    &    & $\de_{10}$ & $2.428\,828\,0568$  \\[0.1cm]
\hline\hline
    \end{tabular}
      \caption{\raggedright Relativistic Ritz fitting parameters for $\ell=0$ and $\ell=1$ HFS states of hydrogen. Note that the numbers are small, since they are to be multiplied
by $10^{-5}$. For states with $2\leq \ell \leq 30$ see Appendix \ref{Appendix_Parameters}.}
        \label{fitting_parameter_table_l01}
  \end{table}

We observe in Table \ref{fitting_parameter_table_l01} and Tables \ref{fitting_parameter_table_l234} - \ref{fitting_parameter_table_l21-30} that 
the behavior of the leading-order defect parameter, $\de_0$, displays a strong dependence on $j$, a weak dependence on $\ell$, and an even weaker dependence on $f$. The scale and variation of $\de_0$ can be understood from the following simplified non-relativistic analysis with fine- and hyperfine structure corrections included. In the $m_e/m_p\to 0$ limit, we can approximate 
\begin{eqnarray}
\f{E}{h}&\simeq&-\f{cR_\infty}{\(n-\de_0\)^2}\notag\\
&\simeq&-cR_\infty\(\f{1}{n^2}  +\f{2}{n^3}\de_0\)\,,
\end{eqnarray}
where in the second line we have assumed $\de_0/n\ll1$, which is verified below. It is well known that fine-structure effects contribute to the energy levels a $j$-dependent term that scales\footnote{The fine-structure correction that scales as $n^{-4}$ is a relativistic kinetic energy correction that is already contained within the relativistic Ritz model -- see equation \eqref{Rel_Ritz_series}.} as $n^{-3}$,
\begin{equation}
\Delta E^\text{(FS)}=-\f{cR_\infty}{n^3}\f{\a^2}{j+1/2} + \dots\,,
\end{equation}
so we should expect that
\begin{equation}
\de_0\simeq \frac{\a^2}{2j+1}\,.
\end{equation}
Hyperfine structure effects contribute to the energy a leading term (see, e.g., \cite{horbatsch2016tabulation}) that is approximately
\begin{equation}
\Delta E^{\text{(HFS)}}\simeq
\begin{cases}
1.42\, \text{GHz}\times\f{\(f-\f{3}{4}\)}{n^3}~&(\ell=0)\\
0.53 \, \text{GHz}\times \f{f(f+1)-j(j+1)-\f{3}{4}}{n^3(2\ell+1)j(j+1)}~&(\ell\neq1)\,,
\end{cases}
\end{equation}
which means that we should expect deviations in the leading order defect due to HFS effects (at fixed $\ell$ and $j$) that are approximately
\begin{equation}
\Delta\(\de_0\)_\text{HFS}\simeq 
\begin{cases}
2.2\times10^{-7}~~&(\ell=0) \\
8.1\times10^{-8}\times j^{-2}~~&(\ell\gg1)\,.
\end{cases}
\end{equation}
This accounts for the approximate differences between the $\de_0$ as seen in Table \ref{fitting_parameter_table_l01} as well as for the rest of the angular momentum channels, up to $\ell=30$, listed in Tables \ref{fitting_parameter_table_l234} - \ref{fitting_parameter_table_l21-30} in Appendix \ref{Appendix_Parameters}. 

We should, however, point out that the precise values for the defect parameters depend somewhat on which QED-predicted levels are used for the fit. For the states of low-lying $\ell$ we have chosen $n_\text{max}=15$, but as an example we reconsider the  $S^{(f=0)}_{j=1/2}$ states by fitting to levels up to $n_\text{max}=16$. A comparison of the parameters between the $n_\text{max}=15$ and $n_\text{max}=16$ fits are shown in Table \ref{compare_fit_parameters_S_12_0_up_to_16}. Minor changes in $\de_0$ are observed, but more substantial changes are seen for the higher-order parameters. Nevertheless, either set of parameters could be used to reproduce the QED-predicted energy levels at a comparable level of accuracy.

  \begin{table}[h]
    \centering
    \begin{tabular}{c r r }
\hline\hline
Parameter & $n_\text{max}=15$ ($\times10^{-5}$) & $n_\text{max}=16$ ($\times10^{-5}$)  \\ \hline
$\de_0$ & $2.5502100611$ & $2.5502099872$  \\
 $\de_2$ & $0.0083755621$ &  $0.0083855869$  \\
$\de_4$ & $-0.0316626562$ & $-0.0320835870$  \\
$\de_6$ & $0.2714656639$ & $0.2786622804$  \\
$\de_8$ & $-1.4934453489$ & $-1.5454747300$  \\
$\de_{10}$ & $3.6057260079$ & $3.7472037620$  \\
$\de_{12}$ & $-2.3560904800$ & $-2.4523221495$  \\
\hline\hline
    \end{tabular}
      \caption{\raggedright Comparison of fitting parameters for $S^{(f=0)}_{j=1/2}$ states between the $n_\text{max}=15$ and $n_\text{max}=16$ fit.}
        \label{compare_fit_parameters_S_12_0_up_to_16}
  \end{table}

Some comments on this procedure are warranted. The defect parameters, $\de_i$, are perhaps best viewed as parameters of a particular fitting function, which is not unique, applied to a particular set of input data, which also is not unique. In fact, there are strong correlations between the parameters; see Table \ref{table:Correlation_Matrix} for the correlation matrix between defect parameters for the $S^{(f=0)}_{j=1/2}$ fit. Therefore, these parameters should not be viewed as fundamental, but a given set of them have a practical use in  reproducing theoretical energy levels without having to use the QED theory directly. When using these parameters, only  the values from a single fit should be used. Furthermore, all reported digits of the parameters up to an absolute precision of $10^{-15}$ should conservatively be used to reproduce the levels below the theoretical uncertainty \eqref{QED_level_unc}. 
\begin{widetext}

\begin{table}[!htb]
\centering
\caption{Correlation Matrix for the $S^{(f=0)}_{j=1/2}$ fit. }
\begin{tabular}{c | r r r r r r r}  \hline\hline
 & $\de_0$ & $\de_2$ & $\de_4$ & $\de_6$ & $\de_8$ & $\de_{10}$ & $\de_{12}$\\  \hline
$\de_0$ & 1.00000 & -0.94504 & 0.88620 & -0.84558 & 0.82121 & -0.80876 & 0.80445 \\
$\de_2$ &-0.94504 & 1.00000 & -0.98466 & 0.96259 & -0.94653 & 0.93770 & -0.93456 \\
$\de_4$ &0.88620 & -0.98466 & 1.00000 & -0.99480 & 0.98747 & -0.98274 & 0.98097 \\
$\de_6$ &-0.84558 & 0.96259 & -0.99480 & 1.00000 & -0.99838 & 0.99641 & -0.99557 \\
$\de_8$ &0.82121 & -0.94653 & 0.98747 & -0.99838 & 1.00000 & -0.99961 & 0.99930 \\
$\de_{10}$  &-0.80876 & 0.93770 & -0.98274 & 0.99641 & -0.99961 & 1.00000 & -0.99996 \\
$\de_{12}$ & 0.80445 & -0.93456 & 0.98097 & -0.99557 & 0.99930 & -0.99996 & 1.00000 \\
\end{tabular}
\label{table:Correlation_Matrix}
\end{table}
\end{widetext}

\ssec{Comparison with experiments}

As an example application of these fits, in Table \ref{QED_meas_discrep_table} we provide a selection of recently measured hydrogen transition frequencies and their corresponding theory predictions using equation \eqref{Rel_Ritz_series}. Weighting by the number of states, the hyperfine centroid is defined as
\begin{equation}
E_{n\ell j}^\text{centroid}=\f{\sum_f (2f+1) E_{n\ell j f}}{\sum_f (2f+1)}
\end{equation}
and the fine-structure centroid is defined as
\begin{equation}
E_{n\ell}^\text{centroid}=\f{\sum_j (2j+1) E_{n\ell j}^\text{centroid}}{\sum_j (2j+1)}\,.
\end{equation}
Following the same rationale leading to equation \eqref{QED_level_unc}, the theoretical error for any given transition is
\begin{multline}\label{QED_transl_unc}
\de(\nu \(n_i\ell_i\to n_f\ell_f\))=\\
\abs{     1.9 \,\text{kHz}\(\f{\de_{\ell_f,0}}{n_f^3} - \f{\de_{\ell_i,0}}{n_i^3}\)    - 2.2 \,\text{kHz}\(\f{1}{n_f^2} - \f{1}{n_i^2} \)    }\,,
\end{multline}
whereas the shift in a transition due to a shift in the proton radius can be easily computed using \eqref{Elevel_change_due_to_rp}.

In some cases the measurement and theory (columns 2 and 5 of Table \ref{QED_meas_discrep_table}) disagree. However, the sums of values in columns 5 and 6 are in good agreement with the measured values in column 2, which confirms that these disagreements are still well characterized by shifts in the proton radius.

\begin{widetext}

  \begin{table}[!htb]
    \centering
    \begin{tabular}{l  r r  c r c}
\hline\hline
      Interval~~~~~~~~~~~~~~~~~~~~~~~~~~~~&Measurement [kHz]& Inferred $r_p$ [fm] & Ref. & This work  [kHz] &  $\Delta (\nu)_{r_p}$ [kHz]  \\ \hline
  $\nu\(1S_{1/2}^{(f=1)}\to3S_{1/2}^{(f=1)}\)$ &  2\,922\,742\,936\,722.4(2.6)\,~& $0.877(13)$~\, & \cite{fleurbaey2018new} & $2\,922\,742\,936\,715.3(1) $ & $+ 7.1$  \\ 
          & 2\,922\,742\,936\,716.72(72)& $0.8482(38)$ &  \cite{grinin2020two} & $2\,922\,742\,936\,715.3(1) $ & $+1.4$  \\[0.2cm]
  $\nu\(2S_{1/2}^{(f=0)}\to2P_{1/2}^{(f=1)}\)$ &  909\,871.7(3.2)\,~& $0.833(10)$~\, & \cite{Bezginov:2019mdi} & $909\,874.1(2)$ & $- 2.6$  \\[0.2cm]
                $\nu\(2S_{}\to4P\)_\text{FS centroid}$ &  616\,520\,931\,626.8(3.3)\,~& $0.8335(95)$ & \cite{beyer2017rydberg} & $616\,520\,931\,628.6(2)$ & $-1.9$  \\[0.2cm]
  $\nu\(2S_{1/2}\to8D_{5/2}\)_\text{HFS centroid}$ &  770\,649\,561\,570.9(2.0)\,~& $0.8584(51)$ & \cite{Brandt:2021yor} & $770\,649\,561\,564.0(3)$ & $+ 6.8$  \\[0.2cm] 
\hline\hline
    \end{tabular}
      \caption{\raggedright A selection of recently measured frequency (energy) intervals of hydrogen and the proton radius values inferred from them in columns 2 and 3, respectively; in column 5 are the bound-state QED predictions using the fitting formula \eqref{Rel_Ritz_series} and defect parameters in Tables \ref{fitting_parameter_table_l01} and \ref{fitting_parameter_table_l234}, which are based on $r_p=0.84087(39)\,\text{fm}$; in column 6 are the proton radius corrections to column 5 using the proton radius from column 3, according to equation \eqref{Elevel_change_due_to_rp}. }
        \label{QED_meas_discrep_table}
  \end{table}
  
\end{widetext}

\sec{Discussion}

Here we have presented a simple fitting formula and parameters, equation \eqref{Rel_Ritz_series} and  Tables \ref{fitting_parameter_table_l01},  and \ref{fitting_parameter_table_l234} through \ref{fitting_parameter_table_l21-30}, that are sufficient to reproduce all hyperfine energy levels of hydrogen up to $\ell=30$. The theoretical uncertainty of any level is given by equation \eqref{QED_level_unc} and additional systematic shifts in those levels due to a proton radius that differs from the one determined by Antognini et al. \cite{Antognini:1900ns} is parameterized in equation \eqref{Elevel_change_due_to_rp}. 

\begin{center}
{\bf Acknowledgements}
\end{center}

We greatly appreciate the initial suggestion of Eric Hessels to pursue this work.

\bibliographystyle{apsrev}

\bibliography{RRitz}

\newpage

\appendix
\sec{Fit parameters for $\ell=2$ through $\ell=30$}\label{Appendix_Parameters}
  \begin{table}[!h]
    \centering
    \begin{tabular}{c  c c c r}
\hline\hline
      \multicolumn{1}{c}
~ $\ell~$ &~$j$~&~$f$~& $\de_i$ & Value$/10^{-5}$  \\ \hline
$2$ &  $\f{3}{2}$ & $1$  & $\de_0$ & $1.332\,822\,5338$  \\
&    &    & $\de_2$ & $0.001\,348\,0751$  \\
&    &    & $\de_4$ & $-0.015\,166\,1825$  \\
&    &    & $\de_6$ & $0.142\,299\,3047$  \\
&    &    & $\de_8$ & $-0.533\,020\,4369$  \\[0.1cm]
$2$ &  $\f{3}{2}$ & $2$  & $\de_0$ & $1.331\,095\,4021$  \\
&    &    & $\de_2$ & $0.001\,348\,1369$  \\
&    &    & $\de_4$ & $-0.015\,166\,1916$  \\
&    &    & $\de_6$ & $0.142\,299\,4035$  \\
&    &    & $\de_8$ & $-0.533\,020\,8355$  \\[0.1cm]
$2$ &  $\f{5}{2}$ & $2$  & $\de_0$ & $0.887\,597\,5544$  \\
&    &    & $\de_2$ & $0.001\,348\,2663$  \\
&    &    & $\de_4$ & $-0.015\,171\,2667$  \\
&    &    & $\de_6$ & $0.142\,369\,4195$  \\
&    &    & $\de_8$ & $-0.533\,332\,7777$  \\[0.1cm]
$2$ &  $\f{5}{2}$ & $3$  & $\de_0$ & $0.886\,487\,6120$  \\
&    &    & $\de_2$ & $0.001\,348\,2916$  \\
&    &    & $\de_4$ & $-0.015\,171\,2659$  \\
&    &    & $\de_6$ & $0.142\,369\,4111$  \\
&    &    & $\de_8$ & $-0.533\,332\,7469$  \\[0.1cm]
$3$ &  $\f{5}{2}$ & $2$  & $\de_0$ & $0.888\,222\,9673$  \\
&    &    & $\de_2$ & $0.000\,781\,8922$  \\
&    &    & $\de_4$ & $-0.008\,421\,7392$  \\
&    &    & $\de_6$ & $0.053\,242\,8585$  \\[0.1cm]
$3$ &  $\f{5}{2}$ & $3$  & $\de_0$ & $0.887\,429\,9452$  \\
&    &    & $\de_2$ & $0.000\,781\,9176$  \\
&    &    & $\de_4$ & $-0.008\,421\,7409$  \\
&    &    & $\de_6$ & $0.053\,242\,8702$  \\[0.1cm]
$3$ &  $\f{7}{2}$ & $3$  & $\de_0$ & $0.665\,695\,0501$  \\
&    &    & $\de_2$ & $0.000\,781\,9366$  \\
&    &    & $\de_4$ & $-0.008\,422\,7940$  \\
&    &    & $\de_6$ & $0.053\,250\,0463$  \\[0.1cm]
$3$ &  $\f{7}{2}$ & $4$  & $\de_0$ & $0.665\,107\,7533$  \\
&    &    & $\de_2$ & $0.000\,781\,9505$  \\
&    &    & $\de_4$ & $-0.008\,422\,7940$  \\
&    &    & $\de_6$ & $0.053\,250\,0464$  \\[0.1cm]
$4$ &  $\f{7}{2}$ & $3$  & $\de_0$ & $0.666\,044\,8195$  \\
&    &    & $\de_2$ & $0.000\,462\,1199$  \\
&    &    & $\de_4$ & $-0.003\,176\,3090$  \\[0.1cm]
$4$ &  $\f{7}{2}$ & $4$  & $\de_0$ & $0.665\,587\,9476$  \\
&    &    & $\de_2$ & $0.000\,462\,1338$  \\
&    &    & $\de_4$ & $-0.003\,176\,3094$  \\[0.1cm]
$4$ &  $\f{9}{2}$ & $4$  & $\de_0$ & $0.532\,550\,4600$  \\
&    &    & $\de_2$ & $0.000\,462\,1300$  \\
&    &    & $\de_4$ & $-0.003\,176\,5643$  \\[0.1cm]
$4$ &  $\f{9}{2}$ & $5$  & $\de_0$ & $0.532\,187\,0988$  \\
&    &    & $\de_2$ & $0.000\,462\,1388$  \\
&    &    & $\de_4$ & $-0.003\,176\,5643$  \\[0.1cm]
\hline\hline
    \end{tabular}
      \caption{\raggedright Same as Table \ref{fitting_parameter_table_l01}, but for $\ell=2-4$ channels.}
        \label{fitting_parameter_table_l234}
  \end{table}

    \begin{table}[!h]%
    \centering
    \begin{tabular}{c  c c c r}
\hline\hline
      \multicolumn{1}{c}
~ $\ell~$ &~$j$~&~$f$~& $\de_i$ & Value$/10^{-6}$  \\ \hline
$5$ &  $\f{9}{2}$ & $4$  & $\de_0$ & $5.327\,754\,771$  \\
&    &    & $\de_2$ & $0.003\,595\,064$  \\
&    &    & $\de_4$ & $-0.033\,558\,712$  \\[0.1cm]
$5$ &  $\f{9}{2}$ & $5$  & $\de_0$ & $5.324\,781\,378$  \\
&    &    & $\de_2$ & $0.003\,595\,152$  \\
&    &    & $\de_4$ & $-0.033\,558\,714$  \\[0.1cm]
$5$ &  $\f{11}{2}$ & $5$  & $\de_0$ & $4.437\,876\,418$  \\
&    &    & $\de_2$ & $0.003\,595\,111$  \\
&    &    & $\de_4$ & $-0.033\,560\,844$  \\[0.1cm]
$5$ &  $\f{11}{2}$ & $6$  & $\de_0$ & $4.435\,406\,543$  \\
&    &    & $\de_2$ & $0.003\,595\,172$  \\
&    &    & $\de_4$ & $-0.033\,560\,844$  \\[0.1cm]
$6$ &  $\f{11}{2}$ & $5$  & $\de_0$ & $4.439\,458\,289$  \\
&    &    & $\de_2$ & $0.002\,004\,614$  \\[0.1cm]
$6$ &  $\f{11}{2}$ & $6$  & $\de_0$ & $4.437\,368\,141$  \\
&    &    & $\de_2$ & $0.002\,004\,674$  \\[0.1cm]
$6$ &  $\f{13}{2}$ & $6$  & $\de_0$ & $3.803\,869\,369$  \\
&    &    & $\de_2$ & $0.002\,004\,589$  \\[0.1cm]
$6$ &  $\f{13}{2}$ & $7$  & $\de_0$ & $3.802\,081\,331$  \\
&    &    & $\de_2$ & $0.002\,004\,634$  \\[0.1cm]
$7$ &  $\f{13}{2}$ & $6$  & $\de_0$ & $3.805\,036\,168$  \\
&    &    & $\de_2$ & $0.001\,679\,042$  \\[0.1cm]
$7$ &  $\f{13}{2}$ & $7$  & $\de_0$ & $3.803\,486\,374$  \\
&    &    & $\de_2$ & $0.001\,679\,086$  \\[0.1cm]
$7$ &  $\f{15}{2}$ & $7$  & $\de_0$ & $3.328\,364\,540$  \\
&    &    & $\de_2$ & $0.001\,679\,019$  \\[0.1cm]
$7$ &  $\f{15}{2}$ & $8$  & $\de_0$ & $3.327\,010\,229$  \\
&    &    & $\de_2$ & $0.001\,679\,053$  \\[0.1cm]
$8$ &  $\f{15}{2}$ & $7$  & $\de_0$ & $3.329\,264\,456$  \\
&    &    & $\de_2$ & $0.001\,433\,670$  \\[0.1cm]
$8$ &  $\f{15}{2}$ & $8$  & $\de_0$ & $3.328\,069\,368$  \\
&    &    & $\de_2$ & $0.001\,433\,704$  \\[0.1cm]
$8$ &  $\f{17}{2}$ & $8$  & $\de_0$ & $2.958\,531\,330$  \\
&    &    & $\de_2$ & $0.001\,433\,650$  \\[0.1cm]
$8$ &  $\f{17}{2}$ & $9$  & $\de_0$ & $2.957\,469\,992$  \\
&    &    & $\de_2$ & $0.001\,433\,677$  \\[0.1cm]
$9$ &  $\f{17}{2}$ & $8$  & $\de_0$ & $2.959\,246\,785$  \\
&    &    & $\de_2$ & $0.001\,242\,839$  \\[0.1cm]
$9$ &  $\f{17}{2}$ & $9$  & $\de_0$ & $2.958\,297\,092$  \\
&    &    & $\de_2$ & $0.001\,242\,865$  \\[0.1cm]
$9$ &  $\f{19}{2}$ & $9$  & $\de_0$ & $2.662\,667\,315$  \\
&    &    & $\de_2$ & $0.001\,242\,821$  \\[0.1cm]
$9$ &  $\f{19}{2}$ & $10$  & $\de_0$ & $2.661\,813\,158$  \\
&    &    & $\de_2$ & $0.001\,242\,843$  \\[0.1cm]
$10$ &  $\f{19}{2}$ & $9$  & $\de_0$ & $2.663\,256\,892$  \\[0.1cm]
$10$ &  $\f{19}{2}$ & $10$  & $\de_0$ & $2.662\,484\,028$  \\[0.1cm]
$10$ &  $\f{21}{2}$ & $10$  & $\de_0$ & $2.420\,605\,509$  \\[0.1cm]
$10$ &  $\f{21}{2}$ & $11$  & $\de_0$ & $2.419\,903\,259$  \\[0.1cm]
\hline\hline
    \end{tabular}
      \caption{\raggedright Same as Table \ref{fitting_parameter_table_l01}, but for $\ell=5-10$ channels.}
        \label{fitting_parameter_table_5-10}
  \end{table}
  
      \begin{table}[!h]
    \centering
    \begin{tabular}{c c c r}
\hline\hline
~ $\ell~$ &~$j$~&~$f$~&~~~$\de_0/10^{-6}$~~  \\ \hline
$11$ &  $\f{21}{2}$ & $10$  & $2.421\,087\,670$  \\[0.1cm]
$11$ &  $\f{21}{2}$ & $11$  & $2.420\,446\,444$  \\[0.1cm]
$11$ &  $\f{23}{2}$ & $11$  & $2.218\,881\,255$  \\[0.1cm]
$11$ &  $\f{23}{2}$ & $12$  & $2.218\,293\,695$  \\[0.1cm]
$12$ &  $\f{23}{2}$ & $11$  & $2.219\,288\,139$  \\[0.1cm]
$12$ &  $\f{23}{2}$ & $12$  & $2.218\,747\,551$  \\[0.1cm]
$12$ &  $\f{25}{2}$ & $12$  & $2.048\,192\,551$  \\[0.1cm]
$12$ &  $\f{25}{2}$ & $13$  & $2.047\,693\,703$  \\[0.1cm]
$13$ &  $\f{25}{2}$ & $12$  & $2.048\,540\,521$  \\[0.1cm]
$13$ &  $\f{25}{2}$ & $13$  & $2.048\,078\,600$  \\[0.1cm]
$13$ &  $\f{27}{2}$ & $13$  & $1.901\,888\,706$  \\[0.1cm]
$13$ &  $\f{27}{2}$ & $14$  & $1.901\,459\,890$  \\[0.1cm]
$14$ &  $\f{27}{2}$ & $13$  & $1.902\,189\,702$  \\[0.1cm]
$14$ &  $\f{27}{2}$ & $14$  & $1.901\,790\,439$  \\[0.1cm]
$14$ &  $\f{29}{2}$ & $14$  & $1.775\,092\,605$  \\[0.1cm]
$14$ &  $\f{29}{2}$ & $15$  & $1.774\,720\,039$  \\[0.1cm]
$15$ &  $\f{29}{2}$ & $14$  & $1.775\,355\,379$  \\[0.1cm]
$15$ &  $\f{29}{2}$ & $15$  & $1.775\,006\,833$  \\[0.1cm]
$15$ &  $\f{31}{2}$ & $15$  & $1.664\,146\,281$  \\[0.1cm]
$15$ &  $\f{31}{2}$ & $16$  & $1.663\,819\,578$  \\[0.1cm]
$16$ &  $\f{31}{2}$ & $15$  & $1.664\,377\,841$  \\[0.1cm]
$16$ &  $\f{31}{2}$ & $16$  & $1.664\,070\,924$  \\[0.1cm]
$16$ &  $\f{33}{2}$ & $16$  & $1.566\,252\,829$  \\[0.1cm]
$16$ &  $\f{33}{2}$ & $17$  & $1.565\,964\,011$  \\[0.1cm]
$17$ &  $\f{33}{2}$ & $16$  & $1.566\,458\,423$  \\[0.1cm]
$17$ &  $\f{33}{2}$ & $17$  & $1.566\,186\,098$  \\[0.1cm]
$17$ &  $\f{35}{2}$ & $17$  & $1.479\,236\,708$  \\[0.1cm]
$17$ &  $\f{35}{2}$ & $18$  & $1.478\,979\,546$  \\[0.1cm]
$18$ &  $\f{35}{2}$ & $17$  & $1.479\,420\,472$  \\[0.1cm]
$18$ &  $\f{35}{2}$ & $18$  & $1.479\,177\,201$  \\[0.1cm]
$18$ &  $\f{37}{2}$ & $18$  & $1.401\,380\,310$  \\[0.1cm]
$18$ &  $\f{37}{2}$ & $19$  & $1.401\,149\,960$  \\[0.1cm]
$19$ &  $\f{37}{2}$ & $18$  & $1.401\,545\,635$  \\[0.1cm]
$19$ &  $\f{37}{2}$ & $19$  & $1.401\,327\,005$  \\[0.1cm]
$19$ &  $\f{39}{2}$ & $19$  & $1.331\,309\,900$  \\[0.1cm]
$19$ &  $\f{39}{2}$ & $20$  & $1.331\,102\,223$  \\[0.1cm]
$20$ &  $\f{39}{2}$ & $19$  & $1.331\,459\,274$  \\[0.1cm]
$20$ &  $\f{39}{2}$ & $20$  & $1.331\,261\,721$  \\[0.1cm]
$20$ &  $\f{41}{2}$ & $20$  & $1.267\,912\,924$  \\[0.1cm]
$20$ &  $\f{41}{2}$ & $21$  & $1.267\,724\,795$  \\[0.1cm]
\hline\hline
    \end{tabular}
      \caption{\raggedright Same as Table \ref{fitting_parameter_table_l01}, but for $\ell=11-20$ channels.}
        \label{fitting_parameter_table_l11-20}
  \end{table}

        \begin{table}[!h]
    \centering
    \begin{tabular}{c c c r}
\hline\hline
~ $\ell~$ &~$j$~&~$f$~&~~~$\de_0/10^{-6}$~~  \\ \hline
$21$ &  $\f{41}{2}$ & $20$  & $1.268\,048\,616$  \\[0.1cm]
$21$ &  $\f{41}{2}$ & $21$  & $1.267\,869\,232$  \\[0.1cm]
$21$ &  $\f{43}{2}$ & $21$  & $1.210\,279\,427$  \\[0.1cm]
$21$ &  $\f{43}{2}$ & $22$  & $1.210\,108\,210$  \\[0.1cm]
$22$ &  $\f{43}{2}$ & $21$  & $1.210\,403\,234$  \\[0.1cm]
$22$ &  $\f{43}{2}$ & $22$  & $1.210\,239\,622$  \\[0.1cm]
$22$ &  $\f{45}{2}$ & $22$  & $1.157\,657\,634$  \\[0.1cm]
$22$ &  $\f{45}{2}$ & $23$  & $1.157\,501\,147$  \\[0.1cm]
$23$ &  $\f{45}{2}$ & $22$  & $1.157\,771\,053$  \\[0.1cm]
$23$ &  $\f{45}{2}$ & $23$  & $1.157\,621\,220$  \\[0.1cm]
$23$ &  $\f{47}{2}$ & $23$  & $1.109\,421\,071$  \\[0.1cm]
$23$ &  $\f{47}{2}$ & $24$  & $1.109\,277\,491$  \\[0.1cm]
$24$ &  $\f{47}{2}$ & $23$  & $1.109\,525\,356$  \\[0.1cm]
$24$ &  $\f{47}{2}$ & $24$  & $1.109\,387\,632$  \\[0.1cm]
$24$ &  $\f{49}{2}$ & $24$  & $1.065\,043\,500$  \\[0.1cm]
$24$ &  $\f{49}{2}$ & $25$  & $1.064\,911\,294$  \\[0.1cm]
$25$ &  $\f{49}{2}$ & $24$  & $1.065\,139\,712$  \\[0.1cm]
$25$ &  $\f{49}{2}$ & $25$  & $1.065\,012\,686$  \\[0.1cm]
$25$ &  $\f{51}{2}$ & $25$  & $1.024\,079\,646$  \\[0.1cm]
$25$ &  $\f{51}{2}$ & $26$  & $1.023\,957\,513$  \\[0.1cm]
$26$ &  $\f{51}{2}$ & $25$  & $1.024\,168\,686$  \\[0.1cm]
$26$ &  $\f{51}{2}$ & $26$  & $1.024\,051\,159$  \\[0.1cm]
$26$ &  $\f{53}{2}$ & $26$  & $0.986\,150\,199$  \\[0.1cm]
$26$ &  $\f{53}{2}$ & $27$  & $0.986\,037\,031$  \\[0.1cm]
$27$ &  $\f{53}{2}$ & $26$  & $0.986\,232\,841$  \\[0.1cm]
$27$ &  $\f{53}{2}$ & $27$  & $0.986\,123\,786$  \\[0.1cm]
$27$ &  $\f{55}{2}$ & $27$  & $0.950\,930\,040$  \\[0.1cm]
$27$ &  $\f{55}{2}$ & $28$  & $0.950\,824\,884$  \\[0.1cm]
$28$ &  $\f{55}{2}$ & $27$  & $0.951\,006\,950$  \\[0.1cm]
$28$ &  $\f{55}{2}$ & $28$  & $0.950\,905\,481$  \\[0.1cm]
$28$ &  $\f{57}{2}$ & $28$  & $0.918\,138\,893$  \\[0.1cm]
$28$ &  $\f{57}{2}$ & $29$  & $0.918\,040\,928$  \\[0.1cm]
$29$ &  $\f{57}{2}$ & $28$  & $0.918\,210\,647$  \\[0.1cm]
$29$ &  $\f{57}{2}$ & $29$  & $0.918\,116\,000$  \\[0.1cm]
$29$ &  $\f{59}{2}$ & $29$  & $0.887\,533\,853$  \\[0.1cm]
$29$ &  $\f{59}{2}$ & $30$  & $0.887\,442\,365$  \\[0.1cm]
$30$ &  $\f{59}{2}$ & $29$  & $0.887\,600\,953$  \\[0.1cm]
$30$ &  $\f{59}{2}$ & $30$  & $0.887\,512\,463$  \\[0.1cm]
$30$ &  $\f{61}{2}$ & $30$  & $0.858\,903\,359$  \\[0.1cm]
$30$ &  $\f{61}{2}$ & $31$  & $0.858\,817\,727$  \\[0.1cm]
\hline\hline
    \end{tabular}
      \caption{\raggedright Same as Table \ref{fitting_parameter_table_l01}, but for $\ell=21-30$ channels.}
        \label{fitting_parameter_table_l21-30}
  \end{table}

\end{document}